\documentclass{sig-alternate}

\usepackage{amssymb}
\usepackage{listings}
\usepackage{booktabs}
\usepackage{mathtools}
\usepackage{tabularx}
\usepackage{fixltx2e}
\usepackage{flushend}

\usepackage{fancyvrb}
\VerbatimFootnotes

\usepackage{graphicx}
\makeatletter
\def\maxwidth#1{\ifdim\Gin@nat@width>#1 #1\else\Gin@nat@width\fi}
\makeatother

\permission{Copyright is held by the author/owner(s).}
\conferenceinfo{WWW 2015 Companion,}{May 18--22, 2015, Florence, Italy.}
\copyrightetc{ACM \the\acmcopyr}
\crdata{978-1-4503-3473-0/15/05. \\
http://dx.doi.org/10.1145/2740908.2742014}

\begin{document}
\title{Science Bots: \\a Model for the Future of Scientific Computation?}

\numberofauthors{1}
\author{
\alignauthor
Tobias Kuhn\\
\affaddr{Department of Humanities, Social and Political Sciences, ETH Zurich, Switzerland}\\
\email{tokuhn@ethz.ch}}
\maketitle

\begin{abstract}

As a response to the trends of the increasing importance of computational approaches and the accelerating pace in science, I propose in this position paper to establish the concept of ``science bots'' that autonomously perform programmed tasks on input data they encounter and immediately publish the results. We can let such bots participate in a reputation system together with human users, meaning that bots and humans get positive or negative feedback by other participants. Positive reputation given to these bots would also shine on their owners, motivating them to contribute to this system, while negative reputation will allow us to filter out low-quality data, which is inevitable in an open and decentralized system.

\end{abstract}



\section{Introduction}

As datasets become increasingly important in all branches of science, many have proposed methods and tools to publish data  \cite{brase2009coinfo} \cite{parsons2010eos} \cite{bechhofer2010fwcs}.
Nanopublications  \cite{groth2010isu} are an approach to bundle atomic data snippets (in RDF) in small packages together with their provenance and metadata. Such nanopublications can be manually created by scientists and linked to their articles, but they can also be automatically extracted from existing datasets or be directly created by programs that implement scientific methods.

In general, computer programs form a third kind of scientific contribution, besides narrative articles and datasets. While many such programs are openly available, there are no conventions or standards of how to reliably link data to the software that produced it, including the version of the software and the input it received.
Moreover, due to the focus of the scientific life cycle on the publication of articles, scientific software is typically applied only to the data available at the time of writing a paper. It is often not the case that new output data is made public when new input data becomes available.
To tackle these problems, I argue that we can encapsulate certain types of scientific algorithms as small independent agents that take inputs of a given type and produce, for example, nanopublications, and they could do this in a real-time and automatic manner as new input data becomes available.

\section{Background}

I borrow the term ``bot'' from Wikipedia, where bots are applied, for example, to revert edits that are the results of vandalism  \cite{halfaker2012sc} \cite{steiner2014wwwc}.
A prominent example is a bot that has created around 454 000 articles for the Swedish Wikipedia  \cite{guldbrandsson2013wikimedia}.
The fact that bots can be powerful also in a negative sense has become apparent with the rise of botnets  \cite{abu2006sigcomm}, and with the increasing problem of "social bots" that pretend to be humans  \cite{ferrara2014rise}.
I argue here that the power of bots could also be harnessed in a positive way for scientific computation.
In contrast to the agents in the original Semantic Web paper  \cite{bernerslee2003sa}, such bots would not propose or make decisions as a kind of personal assistant, but they would only publish data snippets and they would do that without any further interaction with humans.

In previous work, I showed how the concept of nanopublications can be extended and I mentioned the possible use of bots to create them  \cite{kuhn2013eswc}.
I also presented an approach to attach cryptographic hash values to nanopublication identifiers to make them verifiable and immutable  \cite{kuhn2014eswc}.
Based on that work, I have started to establish a nanopublication server network, with which nanopublications can be published, retrieved, and archived in a reliable, trustworthy, and decentralized manner  \cite{kuhn2014publishing}, which could serve as the basis for the communication for bots.

\section{Approach}

In this position paper, I propose bots as a general concept for scientific computation. For example, a bot could apply text mining to extract relations from the abstracts of the constantly growing PubMed database, another bot could regularly measure the temperature at a given location and publish the results, and yet another one could infer new facts from existing nanopublications by applying specified rules or heuristics (e.g. if disease X is related to gene Y, which is targeted by drug Z then Z might help to treat X).
Importantly, these bots can automatically publish the obtained data without double-checking or direct supervision by their creators, and these data can be made immediately accessible to everybody (including other bots).

In a system that treats bots as first-class citizens, we have to expect that some bots (and humans for that matter) will produce low-quality contributions, and we have to make sure that this does not affect the reliability and trustworthiness of the system. I argue that we can achieve that without introducing a central authority, without making concessions with respect to the openness of the system, and without delaying the publishing of results. We simply need a sufficiently accurate automatic method to discern good contributions from bad ones, which can be achieved by a reputation system.
We can let scientists and bots participate in the same reputation system, where they would increase their reputation by receiving positive feedback by other participants on the usefulness and quality of their contributions. Positive reputation of a bot, in turn, would give credit and reputation to the scientist who created it.

To arrive at a simple exemplary model to explain the approach, we can define a relation "is contributed by", where bots can occur on either side: They are contributions, as they were programmed and created by somebody, but they are also contributors, as they can create new digital entities on their own.
We can define a second type of relation to represent assessments. For the sake of simplicity, we model here only positive assessments and strip them of all granularity and detail, and we can call the resulting relation ``gives positive assessment for''.

\begin{figure}[h!]
\centering
\includegraphics[width=\maxwidth{\columnwidth}]{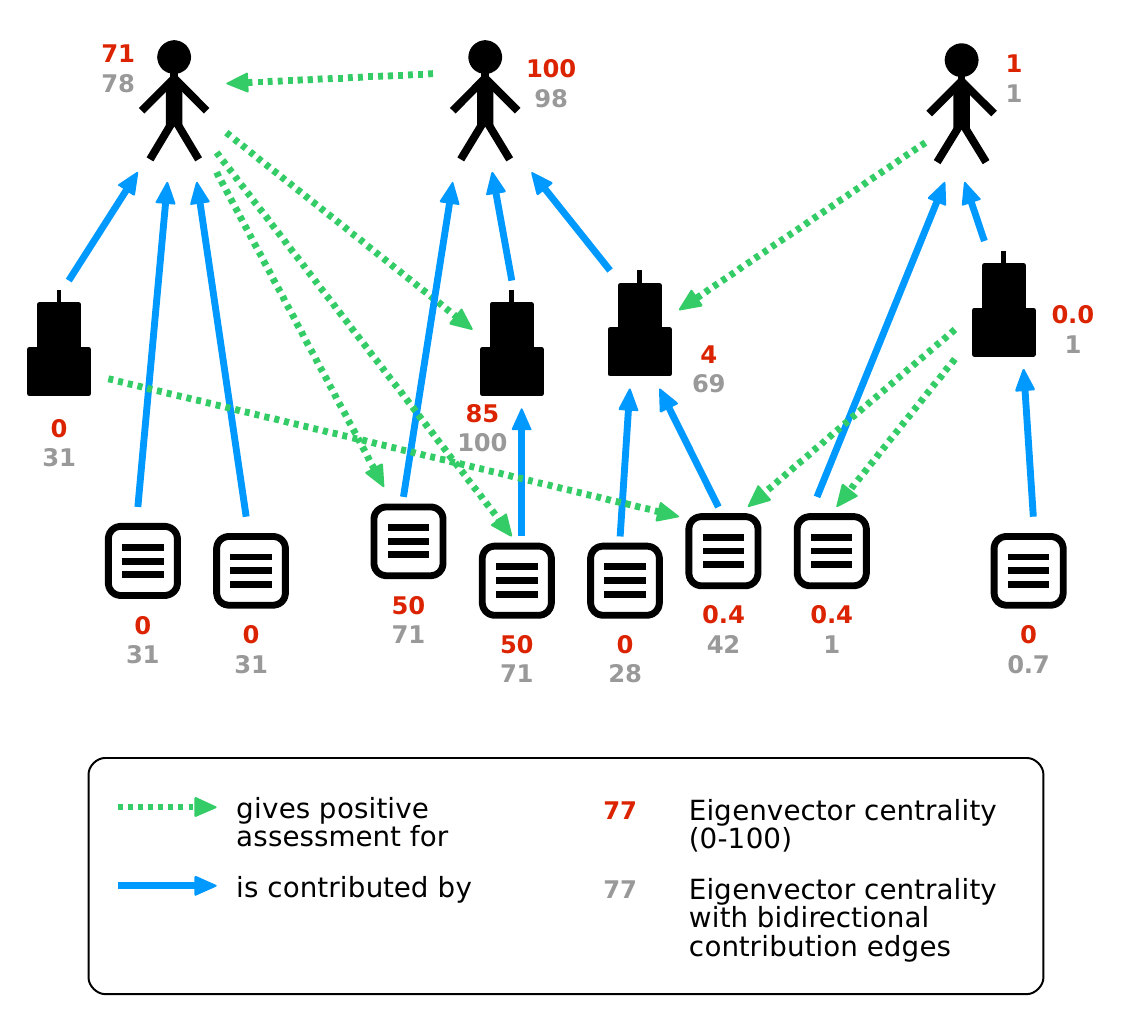}
\caption{
A simple example of a graph of contributors and contributions with edges for creatorship and assessments.
}
\label{figure-graph}
\end{figure}

Fig.~\ref{figure-graph} shows a simple example of such a graph with two kinds of edges, representing creatorship and assessments. To determine the reputation or importance of the nodes, we can in the simplest case treat the two types of edges identically and rank the nodes by applying a network measure such as Eigenvector centrality (which is closely related to Google's PageRank algorithm to rank websites), as shown by the red numbers.
The person at the top-left has a high reputation because he is endorsed by the person in the middle. The latter has a high reputation because her direct and indirect contributions were positively assessed by others (even though she has not received a direct assessment herself). The third person to the right, however, has not contributed anything that was positively assessed by others (only by his own bot), and therefore his reputation is low.
Of course, there are many possible variations and extensions, such as bidirectional contribution edges for the Eigenvector calculation, as indicated by the gray numbers. In general, as one cannot influence {\em incoming} links from the part of the network that is not under one's control, there is no way to efficiently game the system. The scalability of such algorithms in open and decentralized systems is demonstrated by their successful application by search engines and peer-to-peer systems  \cite{kamvar2003www}.

Bots could free scientists from routine tasks and therefore allow them to focus on the interesting questions. Furthermore, this approach could increase the value and appreciation of datasets and software as research products, and give due credit to their creators. With appropriate reputation mechanisms, this can be achieved in a fully open and decentralized environment.

\end{document}